# Colloidal spirals in nematic liquid crystals

Bohdan Senyuk,[a] Manoj B. Pandey,[a] Qingkun Liu,[a] Mykola Tasinkevych[bc] and Ivan I. Smalyukh[*adef]

[a] Department of Physics, University of Colorado, Boulder, CO 80309, USA.
[b] Max-Planck-Institut für Intelligente Systeme, Heisenbergstraße 3, D-70569 Stuttgart, Germany
[c] IV. Institut für Theoretische Physik, Universität Stuttgart, Pfaffenwaldring 57, D-70569 Stuttgart, Germany
[d] Department of Electrical, Computer, and Energy Engineering, University of Colorado, Boulder, CO 80309, USA
[e] Liquid Crystal Materials Research Center and Materials Science and Engineering Program, University of Colorado, Boulder, CO 80309, USA
[f] Renewable and Sustainable Energy Institute, National Renewable Energy Laboratory and University of Colorado, Boulder, CO 80309, USA

*E-mail: ivan.smalyukh@colorado.edu

## Abstract

One of the central experimental efforts in nematic colloids research aims to explore how the interplay between the geometry of particles along with the accompanying nematic director deformations and defects around them can provide a means of guiding particle self-assembly and controlling the structure of particle-induced defects. In this work, we design, fabricate, and disperse low-symmetry colloidal particles with shapes of spirals, double spirals, and triple spirals in a nematic fluid. These spiral-shaped particles, which are controlled by varying their surface functionalization to provide tangential or perpendicular boundary conditions of the nematic molecular alignment, are found inducing director distortions and defect configurations with non-chiral or chiral symmetry. Colloidal particles also exhibit both stable and metastable multiple orientational states in the nematic host, with a large number of director configurations featuring both singular and solitonic nonsingular topological defects accompanying them, which can result in unusual forms of colloidal self-assembly. Our findings directly demonstrate how the symmetry of particle-generated director configurations can be further lowered, or not, as compared to the low point group symmetry of solid micro-inclusions, depending on the nature of induced defects while satisfying topological constraints. We show that achiral colloidal particles can cause chiral symmetry breaking of elastic distortions, which is driven by complex three-dimensional winding of induced topological line defects and solitons.

## Introduction

A plethora of physical phenomena accessible in dispersions of micrometer and submicrometer colloidal particles in liquid crystals (LCs) inspired the development of a rapidly growing branch of soft matter physics known as nematic colloids.[1–18] Gaining a deeper insight into the properties of spontaneous and directed assembly of colloidal particles mimicking naturally occurring systems in atomic and molecular condensed matter can inspire and steer the engineering of artificial functional materials with unusual physical behavior not encountered in nature.[19] Anisotropy of LC properties additionally expands the diversity of colloidal self-assembly. Particles dispersed into LCs induce defects and deformations of a nematic director field $\mathbf{n}(\mathbf{r})$, forming nematic colloidal particles with the inclusions typically surrounded by coronas of elastic distortions of the molecular alignment propagating to large distances, which can be analysed in terms of elastic multipoles in analogy to their electrostatic counterparts.[2,20–25] Two- and three-dimensional crystalline assemblies[6,16] and quasicrystalline tiling[26] were successfully produced using LC elasticity driven interactions between elastic dipoles and quadrupoles around spherical[6,16] and pyramid[26] colloidal particles. Recent studies suggest that $\mathbf{n}(\mathbf{r})$ deformations, induced topological defects, elastic interactions and ensuing self-assembly also strongly depend on the geometric shape and topology of colloidal particles.[8,27–33] Symmetry of $\mathbf{n}(\mathbf{r})$ deformations is dictated by the interplay between the geometrical shape and topology of colloidal inclusions and surface anchoring at their surface. Spherical and other high-symmetry particles can induce relatively low-symmetry deformations of $\mathbf{n}(\mathbf{r})$, which are typically achiral, similar to the colloids themselves. Director field configurations with chiral symmetry were achieved around embedded inclusions, for example, by using spherical particles dispersed in chiral (twisted) LCs,[13,34–36] assemblies of spherical colloids in achiral nematics entangled by disclination lines[37,38] and chiral (often topologically nontrivial) colloidal particles in achiral nematic LCs.[39] Chiral symmetry breaking of director distortions in uniform achiral nematics can be caused not only by the symmetry of the colloidal inclusions but also by tight spatial confinement and strong anisotropy of LC elastic properties as was shown for nematics confined in droplets,[40–42] capillaries[43] or hosting spherical inclusions.[44]

In this article, we study properties of elastic multipoles created in nematic LCs by low-symmetry colloidal particles of the topology of a sphere (genus $g = 0$), but with complex geometrical shape of single, double and triple spirals (to which we will also refer as 1-spirals, 2-

spirals, and 3-spirals, respectively). We show that these spiral-shaped particles can induce director distortions and defect configurations with non-chiral or chiral symmetry, which are controlled by varying surface functionalization of particles to provide tangential or homeotropic boundary conditions for the nematic director. These nematic colloidal particles also exhibit both stable and metastable orientational states with respect to the homogeneous far-field director $\mathbf{n}_0$ and induce a large number of configurations featuring both singular and solitonic nonsingular defects accompanying them, which can result in unusual forms of colloidal self-assembly. Our experimental findings agree with results of numerical modeling based on the minimization of Landau–de Gennes free energy supplemented by surface anchoring terms. Presented results directly demonstrate how complex three-dimensional (3D) winding of topology-dictated line defects can further lower the symmetry of nematic coronas around particles as compared to the low point group symmetry of the particles themselves. We show that this mechanism is also responsible for a chiral symmetry breaking of nematic director distortions generated even by achiral colloidal particles in achiral mesomorphic hosts.

## Materials and experimental techniques

Spiral colloidal particles were fabricated from silica ($SiO_2$) using direct writing laser photolithography. A 90 nm thick aluminium sacrificial layer was sputtered on a silicon wafer first, following by plasma-enhanced chemical vapour deposition of a 1 μm thick layer of silica and spin-coating a layer of photoresist AZ5214 (Clariant AG) on the top. The pattern of spirals was produced first in the photoresist layer by illumination at 405 nm with a direct laser-writing system DWL 66FS (Heidelberg Instruments) and then in the silica layer by inductively coupled plasma etching of uncovered $SiO_2$. The photoresist mask was then removed with acetone, freeing arrays of silica spirals on the top of the supporting aluminium layer (Fig. 1a–c); transparent particles are visible in the reflective mode of optical microscopy mainly due to the scattering of visible light from the edges of particles. During the following preparation stage, the supporting sacrificial layer was removed by wet-etching of aluminium with an aqueous solution of sodium hydroxide (2 wt%) and spiral particles were released from the wafer, while being repeatedly washed, and re-dispersed in deionized water. Single, double and triple spirals with a square (~1×1 μm$^2$) cross-section of 1–3 spiraling arms, respectively, have a lateral dimension of approximately 10, 20 and 23 μm (Fig. 1). To produce a homeotropic alignment boundary conditions for the **n(r)** of the surrounding nematic LCs 5CB (4-cyano-40-pentylbiphenyl from Frinton Laboratories, Inc.) and E7 (EM Industries), these colloidal spirals were treated with an aqueous solution (0.05 wt%) of N,N-dimethyl-N-octadecyl-3-aminopropyl-trimethoxysilyl chloride (DMOAP) and then re-dispersed in methanol. Tangential alignment of **n(r)** at the surface of spirals was achieved by mixing clean untreated silica particles into a negative-dielectric-anisotropy nematic material ZLI-2806 (EM Industries). After the addition of nematic LC and the evaporation of methanol at 70°C overnight, the ensuing nematic dispersion was infiltrated into the cells composed of glass plates separated by glass spacers defining the corresponding thickness of a cell gap. Substrates were treated with DMOAP to achieve a perpendicular $n_0$ or spin-coated with polyimide PI2555 (HD Microsystems) for homogeneous in-plane alignment of $n_0$ enforced by unidirectional rubbing after polyimide cross-linking through baking at 270°C for 1 h. One of the substrates utilized in the cell fabrication was only 150 μm in thickness to minimize spherical aberrations during the imaging and optical manipulation experiments involving high numerical aperture (NA) immersion oil objectives.

Conventional polarizing optical microscopy observations, multi-photon excitation fluorescence polarizing microscopy and optical manipulations were performed using a single multimodal experimental setup built around an inverted Olympus IX81 microscope. A tunable (680–1080 nm) Ti:sapphire oscillator (140 fs, 80 MHz, Chameleon Ultra-II, Coherent) was used for the multi-photon-absorption-based excitation fluorescence polarizing imaging.[45,46] Images of **n(r)** in 5CB and E7 nematics were obtained in the three-photon excitation fluorescence polarizing microscopy (3PEF-PM) mode. This 3PEF-PM imaging in our experiments was realized by the multi-photon-absorption-based excitation of cyanobiphenyl groups of 5CB or E7 molecules by femtosecond laser light at 870 nm; the ensuing fluorescence signal was detected within a spectral range of 387–447 nm by a photomultiplier tube H5784-20 (Hamamatsu).[45] Additionally, we used two-photon excitation fluorescence polarizing imaging (2PEF-PM) to obtain 3D textures of **n(r)** in a nematic ZLI-2806 doped, for this purpose, with a small amount (0.05 wt%) of anisotropic dye *n,n′*-bis(2,5-di-*tert*-butylphenyl)-3,4,9,10-perylenedicarboximide (BTBP, Aldrich), which aligns with transition dipoles of absorption and fluorescence along **n(r)**.[47,48] The dye was excited at 980 nm and its fluorescence was detected within a spectral range of 510–560 nm. Within both imaging approaches, a mirror scanning unit (FV300, Olympus) was used to control an in-plane position of a focused excitation beam and its polarization was varied using a half-wave plate mounted immediately before a 100× (NA = 1.42) oil immersion objective. Optical manipulation of spiral particles was realized with a holographic optical trapping system[45,47] operating at a wavelength of λ = 1064 nm. The same objectives were used for imaging as well as optical manipulations. Optical microscopy textures of a colloidal spiral were recorded with a CCD camera (Flea, PointGrey). We also used an accessory "red plate", which is a full wave retardation plate for the linearly polarized light of a wavelength 530 nm, to determine the direction of **n(r)** in the polarizing microscopy textures. The red plate was inserted before an analyzer with its "slow axis" Z oriented at 45° between crossed polarizers. For relatively thin studied samples, the color of transmitted light in the texture obtained with the red plate is bluish when **n(r)**||Z, yellowish when **n(r)**⊥Z and magenta when **n(r)** is parallel either to a polarizer or analyzer or perpendicular to the plane of view, revealing information about **n(r)** and was used in conjunction with 3PEF-PM imaging to determine **n(r)**-field and defect configurations in our soft matter system.

**Numerical modeling approach**

To theoretically model equilibrium director field configurations around our colloidal spirals, we employ minimization of the Landau–de Gennes (LdG) free energy functional of a tensor field $Q_{ij} = Q_{ji}$, where $i, j = 1,...,3$, which may be written as

$$F = \int_V \left( aQ_{ij}^2 - bQ_{ij}Q_{jk}Q_{ki} + c(Q_{ij}^2)^2 + (L_1/2)\partial_k Q_{ij} \partial_k Q_{ij} + (L_2/2)\partial_j Q_{ij} \partial_k Q_{ik} \right) dV + W \int_{\partial V} f_s ds \quad (1)$$

where summation over repeated indices is assumed; $a$ is assumed to depend linearly on temperature $T$ as $a = a_0(T - T^*)$ with $a_0$ being a material dependent constant and $T^*$ being the supercooling temperature of the isotropic phase; $b$, $c$ are temperature independent material constants. $L_1$ and $L_2$ are phenomenological parameters which may be related to the Frank–Oseen bend, splay and twist elastic constants by using an uniaxial ansatz $Q_{ij} = 3Q_b(n_i n_j - \delta_{ij}/3)/2$, where $Q_b$ is the bulk value of the scalar orientational order parameter, $n_i$ are the Cartesian components of **n(r)**. This gives $K_{11} = K_{33} = 9Q_b^2(L_1 + L_2/2)/2$, and $K_{22} = 9Q_b^2 L_1/2$. The first integral in eqn (1) is taken over the 3D domain of a volume $V$ occupied by a nematic host, whereas the second integral is taken over the surface $\partial V$ (*e.g.*, the surface of colloidal particles) and accounts for non-rigid anchoring surface boundary conditions with an anchoring strength coefficient $W$. The explicit form of the surface free energy density $f_s$ depends on the anchoring type, *e.g.* normal or planar. For the first case we use $f_s^\perp = (Q_{ij} - Q_{ij}^s)^2$ with $Q_{ij}^s = 3Q_b(v_i v_j - \delta_{ij}/3)/2$ being the surface induced value of the tensor order parameter and $\mathbf{v}$ is the outward unit normal vector to the surface. For a planar degenerate anchoring we use $f_s^\parallel = (\tilde{Q}_{ij} - \tilde{Q}_{ij}^\perp)^2 + (\tilde{Q}_{ij}^2 - \frac{3}{2}Q_b^2)^2$, where $\tilde{Q}_{ij} = Q_{ij} + Q_b \delta_{ij}/2$, and $\tilde{Q}_{ij}^\perp = (\delta_{ik} - v_i v_k)\tilde{Q}_{kl}(\delta_{lj} - v_l v_j)$.[49] The free energy in eqn (1) is numerically minimized by using the adaptive finite elements method.[50] The domain $V$ is discretized by using the Quality Tetrahedral Mesh Generator,[51] while the surface of a single spiral particle is triangulated with the help of GNU Triangulated Surface Library.[52] Linear triangular and tetrahedral elements are used and the integration over the elements is performed numerically by using fully symmetric Gaussian quadrature rules.[53–55] The discretized Landau–de Gennes functional is minimized with the help of the INRIA's M1QN3 optimization routine.[56]

The spatially uniaxial nematic phase is thermodynamically stable at $\tau \equiv 24ac/b^2 < 1$ (in calculations $\tau \approx 0.16$), and $Q_b = (b/8c)(1 + \sqrt{1 - 8\tau/9})$. We use $a_0 = 0.044\times10^6$ J K$^{-1}$ m$^{-3}$, $b = 0.816\times10^6$ J m$^{-3}$, $c = 0.45\times10^6$ J m$^{-3}$, $L_1 = 6\times10^{-12}$ J m$^{-1}$, and $L_2 = 12\times10^{-12}$ J m$^{-1}$, which are typical values for 5CB,[57] and $T^* = 307$ K. For these values of the model parameters, the bulk correlation length can be found as $\xi = 2\sqrt{2c(3L_1 + 2L_2)}/b \approx 15$ nm at the isotropic–nematic coexistence at $\tau = 1$.[58] Computer-simulated director configurations induced by the spiral-shaped particles, which correspond to local and global minima of free energy, are visualized and then compared to experimental findings.

## Results and discussion

### Geometry of colloidal inclusions

The symmetry of the director field distortions introduced into the otherwise homogeneous LC by a colloidal particle is dependent on the symmetry of the particle itself, strength of surface boundary conditions in terms of the molecular alignment and also colloidal particle orientation with respect to the far-field director $\mathbf{n}_0$, as well as other factors that we discuss below. In our experiments, each single spiral particle (Fig. 1a and d) is designed to have a shape of the Archimedean spiral, which in polar or cylindrical coordinates can be described as $r = p\theta$, where $r$ and $\theta$ are a radial distance and a polar angle, respectively, and $p$ is a geometric parameter that controls the spacing between successive turns. In our experimental design, two such single spirals, rotated with respect to each other around a normal to their plane $\boldsymbol{a}$ (Fig. 1e) and connected at their outer ends, constitute the 2-spiral particle (Fig. 1b) and are fabricated as a single continuous object. Similarly, three connected single spirals form a triple colloidal spiral (3-spiral, see Fig. 1c). The point group symmetry of a single colloidal spiral object is $C_s$ as the only symmetry operation for this colloidal particle is reflection through a horizontal plane normal to $\boldsymbol{a}$ (Fig. 1e). Symmetry of multi- or m-spirals with m spiral branches increases to $C_{mh}$ because, in addition to mirror reflection through a horizontal plane perpendicular to $\boldsymbol{a}$, they also allow for m-fold rotations around a principal axis $C_m$ parallel to $\boldsymbol{a}$. Therefore, double and triple spirals have the $C_{2h}$ and $C_{3h}$ symmetry (Fig. 1f and g), respectively. All three particles are achiral in the 3D space, which is because of the presence of the mirror symmetry plane coinciding with the particle's midplane.

Similar to other colloidal particles,[2,6,39,59,60] our micrometer-sized spirals induce deformations of director field when introduced into an aligned nematic LC (Fig. 2–7). The homogeneous nematic fluid, a host medium for these spiral-shaped particles, has a rotational symmetry ($D_{\infty h}$) around $\mathbf{n}_0$,[61] which in our experiments can be aligned, for example, along the $z$ axis ($\mathbf{n}_0 = \{0,0,1\}$) in homeotropic cells (Fig. 2) or along different in-plane directions defined by unidirectional rubbing, such as along the $y$ axis in Fig. 3 or $x$ axis in Fig. 7. Hence, although the symmetry of $\mathbf{n}(\mathbf{r})$ distortions induced by colloidal spirals can be different from symmetries of particles themselves, it cannot be higher than the symmetry of the undistorted LC with the ground-state uniform $\mathbf{n}_0$[23–25,61] or that of the colloidal inclusions. One of the goals of this work is to explore how the symmetry of $\mathbf{n}(\mathbf{r})$ configurations relates to the point symmetry of the particles themselves.

**Spirals with homeotropic surface anchoring**
In a nematic LC, colloidal spirals with homeotropic surface anchoring most typically align so that their axis $\boldsymbol{a} \| \mathbf{n}_0$ (Fig. 2–6). Colors in polarizing microscopy textures with a red plate (Fig. 2d–

f) and distribution of intensity in 3PEF-PM textures (Fig. 2j–l) with respect to the polarization of excitation light confirm a homeotropic alignment of LC molecules at the surface of spirals. Due to thermal fluctuations, spiral particles exhibit both translational and rotational diffusion constrained by their orientation with respect to $\mathbf{n}_0$. Their translational diffusion in the plane normal to $\mathbf{n}_0$ and rotational diffusion around $\mathbf{n}_0$ can be observed in homeotropic cells (Fig. 2a). Geometry of the experiment in planar cells (Fig. 3a) allows observation of Brownian motion of colloidal spirals along $\mathbf{n}_0$ and angular fluctuations of a with respect to $\mathbf{n}_0$. Translational diffusion across the homeotropic and planar cells is restricted by the elastic repulsion from the confining walls, which keeps particles in the bulk of the sample (see, for example, Fig. 2m, 3d and 7f), away from confining substrates.

Director fields induced by homeotropic spirals deviates from $\mathbf{n}_0$ most strongly in the spatial locations where the normal $v$ to the spiral's surface is perpendicular to $\mathbf{n}_0$; the director remains roughly intact far from the spiral and at the spiral's faces aligned with their $v \parallel \mathbf{n}_0$. The resulting singular half-integer disclinations wrap the spiral body along the perimeter of faces aligned with $v \perp \mathbf{n}_0$ (normal walls) closely following the perimeter of the spiral object and mimicking the particle's shape (Fig. 2a and p). Such disclinations are the only topologically stable $\pi_1(S^2/Z_2)$ defect lines in the bulk of 3D nematic LC samples.[40,62] Although one can often characterize the details of the local director structure of such half-integer defect lines by winding numbers $k = \pm 1/2$, with the sign determined by the direction in which $\mathbf{n}$ rotates by $\pi$ around its defect core relative to the navigation direction, the disclinations of opposite $k$ can continuously morph one to another in 3D and are topologically similar objects.[62] The outer normal wall of the spiral is encircled by the disclination with the local director structure characterized by $k = -1/2$ but, depending on the way this disclination further propagates in the space between the inner normal walls of successive turns of the spiral, deformations in homeotropic samples can result in the formation of a flat loop (FL) and a jumping loop (JL) of a disclination, as well as a solitonic nonsingular escaped disclination configuration (EC) of $\mathbf{n(r)}$ and different combinations of such defect structures. In the FL configuration, a singular disclination forms the closed loop, which closely follows the perimeter of the spiral along the outer and inner side faces/walls (Fig. 2a, d, g, j, m, p and 4a). Fragments of the closed disclination loop propagating in between two inner walls in JL can jump to the opposite wall (Fig. 2b, e, h, k, n, r, s and 4f). When propagating along the same (Fig. 2s) or opposite (Fig. 2r) wall, these fragments of the disclination line are shifted vertically to avoid each other, localizing nearby opposite edges of the wall rather than at its middle (Fig. 2r and s). This pair of the disclination fragments can propagate along one or both of the inner walls all the way within the interior of the spiral (Fig. 4f) or, in some cases, can jump back and forth between the interior walls and locations in the particle midplane and at its edges (Fig. 2b). The sample segments where line fragments propagate at the same or opposite inner wall are seen as bright birefringent regions filling the entire gap between the inner walls due to locally strongly distorted director oriented away from homeotropic $n_0$ (Fig. 2b and e).

Closed disclination loops are topologically equivalent to hedgehog point defects with integer charges $N_h$.[15,40,62] The topological hedgehog charge of such defect loops around individual spirals in FL and JL configurations can be determined as $N_h = 1$ or $N_h = -1$, because of the nonpolar character of the director (**n**≡-**n**). Decorating the nonpolar n with arrows and transforming it into a vector field allows one to determine the relative signs of the topological hedgehog charges of defects in the multi-defect textures such as the ones observed in our experiments. For example, choosing the outward direction of the decorating vector field at the surface of the spiral particle one can determine the topological hedgehog charge of the disclination loops in our experiments as $N_h = -1$. When the direction of the decorating vector field at the particle surface is reversed to be inward, the sign of the induced topological defect changes to $N_h = 1$. In both cases, the hedgehog charge of the induced defect $N_h = \pm 1$ compensates for the oppositesign topological hedgehog charge of the decorating vector field on the surface of the particle $N_p = \pm\chi/2 = \pm 1$, with the opposite signs of $N_p$ and $N_h$ dependent on the direction of vectorization of **n** on the particle surface. Therefore, the $N_h = \pm 1$ charge of the induced defect is consistent with topological theorems[63] and Euler characteristic $\chi = 2$ (genus $g = 0$) of the spiral-shaped particle homeomorphic to a sphere. In the third type of **n(r)** configuration (EC) (Fig. 2c, f, i, l, o and q), which is the most nontrivial one, two ends of the outer half-integer disclination meet in the region between inner walls and merge into a solitonic "escaped" nonsingular configuration (Fig. 2q). In experiments, we observed mostly the FL and, less frequently, JL configurations; the escaped configuration was observed on a rare occasion. Also, we found that EC is less stable as compared to FL: when locally melted with a laser tweezers and quenched back to the LC phase, EC does not re-appear but rather transforms into either the FL or JL configurations. Experimentally observed and reconstructed **n(r)** corresponding to the FL and JL configurations agree with numerical simulations (Fig. 4a and f). The elastic free energy due to the particle-induced director distortions depends on the particle orientation with respect to the far-field director (Fig. 4k).

The symmetry of the FL and EC of director field distortions around 1-spirals in homeotropic cells is the same as that of the particle itself, *i.e.*, $C_s$. However, LC elasticity-mediated locking of the particles center of masses in the cell midplane as well as locking of the particles normals parallel to $\mathbf{n}_0$ induces effective two-dimensional (2D) chiral symmetry of the $\mathbf{n}(\mathbf{r})$ configuration around the spirals (Fig. 2a and c): their mirror images with respect to a plane containing a cannot be superimposed within the midplane of the cell by translation and rotation operations only in the plane of the cell. Complex 3D propagation of disclinations in the JL configuration modifies the symmetry of resulting $\mathbf{n}(\mathbf{r})$ even more significantly. Because fragments of a disclination loop jump between inner walls (see also result of simulations in Fig. 4f) at different angles and locations with respect to a horizontal mirror plane normal to *a* (Fig. 2r and s), the reflection symmetry with respect to this plane (Fig. 1e) is broken. This lowers the point group symmetry of the *n*(*r*) configuration to $C_1$ (asymmetric) and makes it 3D chiral (Fig. 2k, n, r and s).

Spiral particles in planar cells most commonly align with $\boldsymbol{a}\|\mathbf{n}_0$ (Fig. 3a–d), but the orientation with $\boldsymbol{a}\perp\mathbf{n}_0$ is also possible (Fig. 3e–h). Similar to the homeotropic cell, spirals in both orientations levitate in the bulk of planar LC samples at some distance from substrates determined by the balance between forces of elastic repulsion of spirals from confining substrates and a gravitational force $F_g = \Delta\rho V_s g \approx 0.55$ pN, where $\Delta\rho \approx 1650$ kg m$^{-3}$ is a difference between density of silica and LC, $V_s \approx 34 \times 10^{-18}$ m$^3$ is a volume of a single spiral particle and $g = 9.8$ m s$^{-2}$ is the gravitational acceleration constant.

Results of numerical modeling of orientation of spiral particles with respect to the director are generally consistent with our experimental findings. Numerical calculations of Landau–de Gennes free energy for spirals with a parallel and perpendicular to $\mathbf{n}_0$ show that, although orientation $\boldsymbol{a}\perp\mathbf{n}_0$ are occasionally observed in experiments (Fig. 3e), the orientation $\boldsymbol{a}\|\mathbf{n}_0$ is energetically more preferable (Fig. 4k). Rotation of the spiral particle away from the state with $\boldsymbol{a}\|\mathbf{n}_0$ changes configurations of $\mathbf{n}(\mathbf{r})$ and increases associated free energy (Fig. 4k). Two of many possible topologically nontrivial $\mathbf{n}(\mathbf{r})$ configurations around 1-spirals at the orientation $\boldsymbol{a}\perp\mathbf{n}_0$ (Fig. 3e) are shown in Fig. 4d and i (Fig. 4e and j show the same particles but only with disclination lines and without the details of the director field). In these cases a single disclination loop observed at $a\|n_0$ (Fig. 4a or f) may split into two separate disclination loops, as detected in the calculated defect and director field configuration in Fig. 4e and j. The close inspection of $n(r)$ along these new disclinations reveals that one of them (a longer loop in Fig. 4j for example) continuously morphs along its contour and is topologically charge-neutral[64,65] with $N_{h1} = 0$. Another disclination loop (the shorter loop seen in Fig. 4j), on the other hand, has a topological hedgehog charge $N_{h2} = \pm 1$. Their sum $N_{h1} + N_{h2} = \pm 1$ agrees with the topological constraints. Transformation of $n(r)$ and accompanying disclinations caused by reorientation of spirals relative to $n_0$ can also significantly modify the overall symmetry of resulting nematic coronas and defects around the spirals. The reflection and chiral symmetries of the ensuing particle-induced $n(r)$ can be broken due to specific entanglement of disclinations with the spiral-shaped particle (Fig. 4e). For example, in the calculated texture of the spiral particle with $a \perp n_0$ (Fig. 4d and e), its initial symmetry $C_s$ at $a\|n_0$ (Fig. 4a) is broken down to $C_1$ as the only symmetry operation allowed here is one full rotation around a principal axis $C\|n_0$ (i.e., the identity operation), so that the particle-induced field configuration becomes chiral in the 3D space as well.

Topological features of director fields around double and triple spirals with homeotropic anchoring are similar to those in the case of single spirals. Multi-spirals typically align with their plane normal to $n_0$, so that the half-integer disclination line with the prevailing $k = -1/2$ local structure encircles around their perimeter (Fig. 5) by closely following the side edges of the particles. Disclinations can entangle every branch of multi-spiral particles differently, thus increasing the variety of unique nematic colloidal field configurations and point group symmetries, which can emerge as a result of incorporation of such particles into the aligned nematic LC host. For example, in the case of formation of the FL configurations of $n(r)$ in both branches of a 2-spiral, the resulting nematic colloidal object (particle with induced defects and elastic distortions) has symmetry $C_{2h}$ similar to the symmetry of the colloidal particle itself, but this symmetry can be lowered to $C_s$ in cases if one of the branches has EC and to the chiral $C_1$ if one or both (Fig. 5a–d) branches have the JL configuration of $n(r)$. In planar samples, the orientation of the multi-spirals is observed to be mostly with $a\|n_0$ (Fig. 5e–i), which can be explained by the fact that the elastic energy cost for having their normal a perpendicular to $n_0$ is significantly higher than in the case of single 1-spirals.

When two colloidal particles are relatively close to each other, their regions of distorted

**n**(**r**) overlap causing the total elastic energy of two particles to be dependent on a separation between them. Minimization of this total energy gives rise to long-range anisotropic elastic forces driving particles closer together or further apart, depending on conditions such as the orientation of the inter-particle separation vector relative to **n**$_0$.[2,36,60,66] We have observed that spiral particles, when brought in vicinity of each other using optical tweezers, interact attractively attaching side to side and form pairs in homeotropic cells (Fig. 6). Within such assemblies, interestingly, colloidal particles do not physically touch each other but are kept apart separated by a narrow region of LC with distortions and defects (Fig. 6d and h). The configuration of **n**(**r**) at the outside walls of spirals (Fig. 2p) resembles the quadrupolar "Saturn ring" configuration of **n**(**r**) around spherical particles with homeotropic anchoring, although morphed to follow the geometry of the solid inclusion.[52] Thus for spirals to interact attractively side to side, one would expect that spirals have to either slightly tilt or vertically displace with respect to each other or, if the confinement is tight and spirals are kept strictly in the middle of the thin cell as in our experiments (Fig. 2m), the disclination lines in a contact point have to be slightly deformed and shifted up and down along the side wall to compensate for this tilt angle. Elastic attractions in the homeotropic cells do not show strong anisotropy of interactions with respect to the in plane orientation of the spirals and they can form a variety of configurations (Fig. 6). The potential of side to side attractive elastic interactions between two spirals was estimated using well established methodology[8,36,59,60,66] and is of the order of $1000k_\mathrm{B}T$. Elastic attractive interactions in our experiments were also significantly affected by close vicinity of confining substrates due to screening of elastic distortions by strong confinement (Fig. 2m). Interestingly, provided that details of formed defect and director field configurations in-between the particles stay the same as when the particles are far apart from each other, the symmetry of an assembly formed by a pair of two spiral nematic colloidal particles with the same FL or JL configuration of **n**(**r**) can be increasing from $C_\mathrm{s}$ to the $C_{2h}$ point group symmetry.

**Spirals with tangential surface anchoring**

Spiral particles with tangential surface anchoring typically align with their horizontal mirror symmetry plane tangential to **n**$_0$ and $a \perp \mathbf{n}_0$ (Fig. 7), in agreement with the numerical calculations of the LdG free energy (Fig. 8a). They induce several stable and metastable configurations of **n**(**r**) with different number and location of surface singular point defects called "boojums".[40,61,67]

The azimuthal orientation of end-to-end vectors of single spirals and center-to-end vectors of 2-spirals and 3-spirals in the plane of planar cells (probed *via* the orientation of a line connecting spiral's ends) is found to be essentially random with respect to $\mathbf{n}_0$. However, orientations of these end-to-end and center-to-end vectors correlates with spatial locations of boojums on the particle surfaces, although the number of boojums stays constant and depends on details of particle geometry. In experiments, examination of many particles with identical geometric shapes reveals that the $C_s$ symmetry of 1-spirals with tangential anchoring tends translating to that of the induced $\mathbf{n}(\mathbf{r})$ in the planar LC cells. The configuration of $\mathbf{n}(\mathbf{r})$ around 3-spirals has symmetry $C_s$ and $C_{2h}$ symmetry of double spirals can either be inherited by the induced local director fields/defects around the particle or (more commonly) be lowered to the $C_s$ point group, depending on the location of boojums. The number of induced boojums is always even and depends on number of turns in one spiral and total number of spiraling arms in a colloidal particle. A single spiral particle with the number of turns used in our design is observed to induce[8] boojums (Fig. 7a–c and j). This number of surface point defects is independent of the azimuthal rotation of the particle (Fig. 7j and k). The structure of director distortions and the number of surface point defects around a single spiral (Fig. 7j and k) deduced from the experiments (Fig. 7a) are in a good agreement with results of numerical calculations (Fig. 8b).

The 2D director field $\mathbf{n}_s(\mathbf{r})$ at the interface of the LC and the immersed particle contains 2D defects corresponding to the boojums with integer winding numbers $s$ (Fig. 7j–m). The winding number (or strength) $s$ is determined as a number of times the director $\mathbf{n}_s(\mathbf{r})$ on the surface of the particle makes a full rotation along the line encircling a 2D singularity of the boojum at the LC–particle interface. The sign of $s$ is assigned based on direction of rotation of the director $\mathbf{n}_s(\mathbf{r})$ with respect to the direction of navigation along the line encircling a boojum. In described case boojums with $s = 1$ and $s = -1$ have a radial (Fig. 7l) and hyperbolic (Fig. 7m) distribution of $\mathbf{n}_s(\mathbf{r})$, respectively. Three out of four pairs of boojums have $\mathbf{n}_s(\mathbf{r})$ defects of opposite sign, self-compensating each other, so that the total topological charge of these surface defects (Fig. 7j–m) $\Sigma_i s_i = 2$ satisfies the constraints set by topological theorems.[63]

Interestingly, computer simulations also show that the number of boojums remains constant during rotation of the spiral from energetically most favorable $\boldsymbol{a} \perp \mathbf{n}_0$ (Fig. 8a and b) to energetically most costly $\boldsymbol{a} \| \mathbf{n}_0$ orientations (Fig. 8a and d). During this rotation process boojums only change their location along the perimeter of cross-section of spiral's arm (Fig. 8b–d). As one can see from calculations, minimum of the LdG free energy does not exactly correspond to $a \perp n_0$, but close to this when the angle between $a$ and $n_0$ is $\theta \approx 80°$ (Fig. 8a). This result of numerical modeling, which is somewhat different from experimental findings, can be related to the used approximations, such as the ones related to elastic constants in minimization of free energy. Details of the cross-sectional shape of the spiral's arms and distance between spiral's turns used in the numerical modeling, which are somewhat different from experiments, could also play a role. In particular, a close inspection of simulated textures (Fig. 8b–d) allows for suggesting that the equilibrium orientation (at which the total free energy is at the minimum) would move further closer to orientation $a \perp n_0$ as the area of cross-section of the spiral's arm decreases and the distance between spiral's turns increases. The same effect could be expected to occur when the cross-section shape change from a square-like to a more rounded and eventually circular one.

**Elastic dipoles of spiral-shaped colloidal particles**

Nematic colloids with director field configurations surrounding spherical particles and their long-range interactions are often discussed in terms of the analogy with electric multipoles in electrostatics,[6,20–22,59,60,66,68,69] although complex shape of colloidal particles can also result in formation of elastic multipoles that have no direct electrostatic analogues. Within one approach developed recently, called "nematostatics",[23–25] elastic dipole configurations of director structures around particles with a complex shape can be described in terms of three coefficients of isotropic $d$, anisotropic $D$, and chiral $C$ strengths and a two-component vector ($\gamma$, $\gamma'$). Depending on the number and combination of nonzero coefficients dictating the resulting full symmetry of $n(r)$ around a colloidal particle, nematostatics distinguishes a large number of different elastic dipoles that can be realized.[23–25]

Using symmetry considerations of the nematostatics approach, we can characterize the elastic multipoles formed by director distortions around our spiral-shaped particles. Single colloidal spirals with homeotropic anchoring and the FL or EC structures have $C_s$ symmetry (Fig. 1e, 2p and q), which also can be referred to as $C_{1h}$. This symmetry of the nematic colloidal object allows for a full rotation around an axis perpendicular to $a$ and $n_0$, also laying in the mirror plane, thus forming biaxial elastic dipoles.[25] Biaxial elastic dipoles with $C_{2h}$ and $C_{3h}$ symmetry can be also formed, respectively, by 2- and 3-spirals with either FL or EC of $n(r)$ in their branches. If branches of multi-spirals have the FL and/or EC **n(r)**-configurations, then the resulting biaxial dipole is of $C_{1h}$ symmetry. The presence of the JL **n(r)**-configuration in a 1-spiral or in one of the branches of multi-spirals lowers the full symmetry of the nematic colloidal particle to $C_1$, forming a general chiral elastic dipole around spirals of all types. Interestingly, this chirality is not induced by the azimuthal anchoring with the helicoidal alignment of easy axis at the surface of particles, as in the case of chiral dipoles discussed in ref. 25, but rather by the nontrivial way in which singular topological defects wind around the colloidal spirals. The director field configurations around 1-, 2-, 3- and other multi-spirals with tangential anchoring boundary conditions (Fig. 7) allow for a full rotation around axis parallel to $\mathbf{n}_0$ and have one horizontal mirror plane thus forming nonchiral biaxial elastic dipoles in the plane of the cell.

## Conclusions

We have described realization and properties of elastic multipoles created in the nematic LCs by low-symmetry colloidal particles with shape of single, double and triple spirals. We have shown that these spiral-shaped particles induce director distortions and defect configurations with non-chiral or chiral symmetry, which can be controlled by varying surface functionalization of particles to provide tangential or homeotropic boundary conditions for the nematic molecular alignment at particle surfaces. These colloidal particles also exhibit both stable and metastable orientational states with respect to the far-field director, with a large number of configurations featuring both singular and solitonic nonsingular defects accompanying them, which can be utilized in enriching colloidal self-assembly. Our experimental observations are also in a good agreement with numerical calculations of the director structures around spiral-shaped particles. Our findings directly demonstrate how the symmetry of nematic colloids can be further lowered, or not, as compared to the low point group symmetry of solid particles, depending on the nature of topology-dictated defects. We also show that dispersion of achiral colloidal particles in achiral nematic fluid hosts can lead to chiral symmetry breaking of a resulting nematic director field driven by symmetry breaking winding of singular topological defect lines and nonsingular solitons around these particles.

## Acknowledgements

We acknowledge discussions with A. Martinez, T. Lee, P. Ackerman and P. Chen. This research

was supported by the US National Science Foundation Grant DMR-1410735 (B.S., M.B.P., Q.L, and I.I.S.). M.B.P. acknowledges support of the Indian Government through the DST-BOYSCAST Fellowship program. I.I.S. acknowledges the hospitality of the Isaac Newton Institute's program "Mathematics of Liquid Crystals" and Max-Planck-Institut für Intelligente Systeme in Stuttgart, Germany during long-term visits within which parts of this work were initiated and done, respectively.

**Figures**

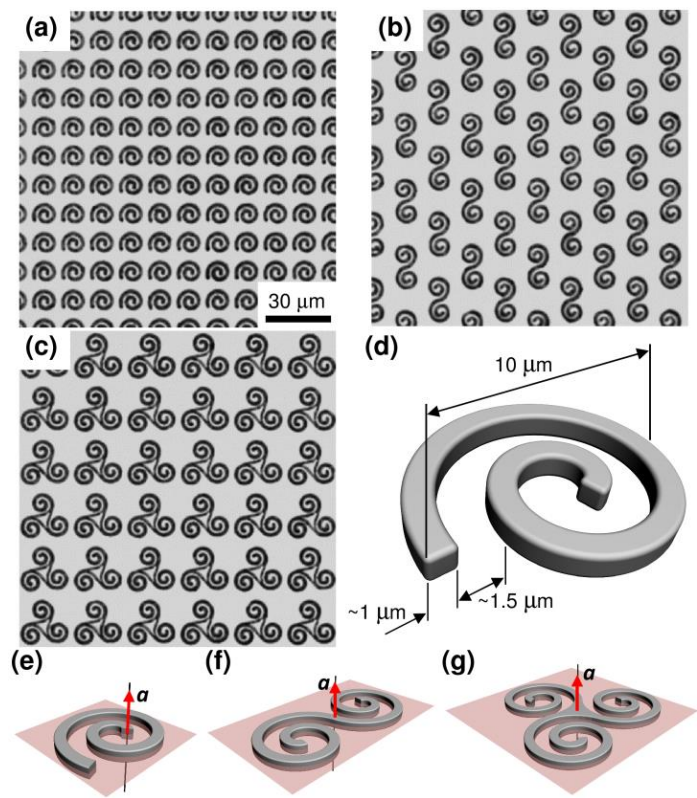

**Fig. 1** Micrographs of arrays of (a) single, (b) double and (c) triple spiral silica colloidal particles on an aluminum layer obtained in the reflective mode of an optical microscopy. (d) Schematic and dimensions of a single spiral particle. (e–g) The corresponding schematics showing the symmetry of spiral particles.

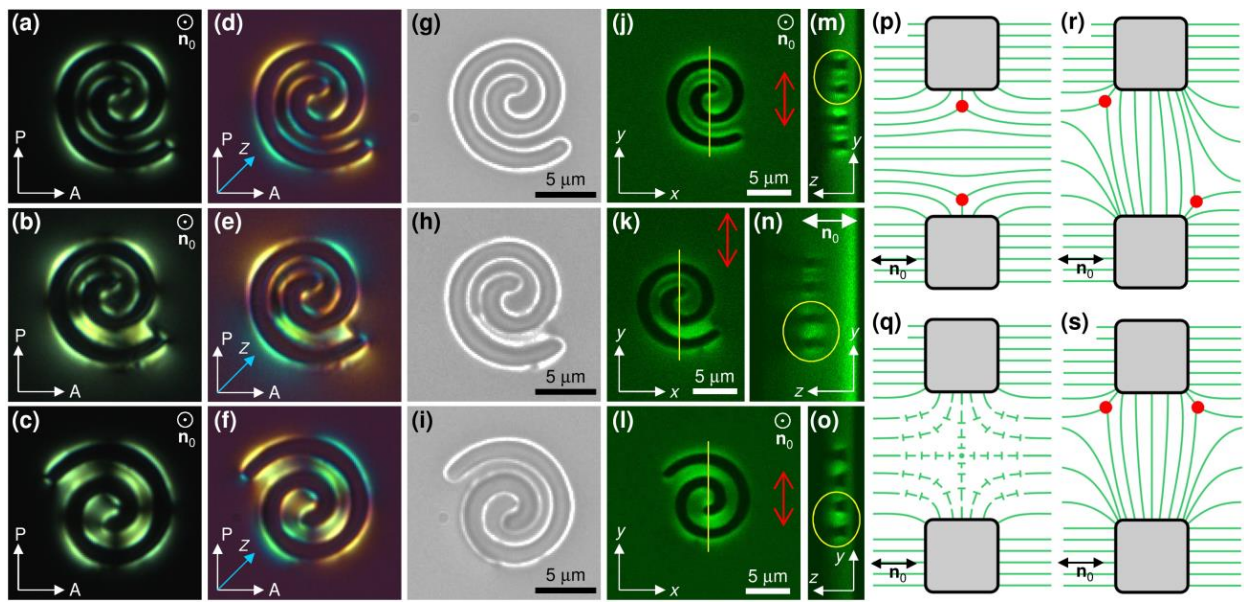

**Fig. 2** (a–o) Micrographs of singular spiral particles with homeotropic anchoring in a homeotropic nematic liquid crystal cell obtained with (a–f) polarizing, (g–i) bright field and (j–o) 3PEF-PM optical microscopies. *P*, *A* and *Z* respectively mark a direction of the crossed polarizer and analyzer and a slow axis of a red plate (blueish color in the texture corresponds to **n**(**r**)∥*Z* and yellowish to **n**(**r**)⊥*Z*). (m–o) 3PEF-PM cross-sectional *zy* images obtained along the bright yellow lines in (j–l). Red double arrows in (j–l) show a polarization of the excitation light. A circle with a central dot and double white and black arrows show a direction of **n**$_0$. (p–s) Schematic diagrams showing **n**(**r**) (green lines) around a fragment of a spiral particle respectively encircled in (m, o and n). Red filled circles show disclinations. Nails in (q) show **n**(**r**) escaping out of a plane of image.

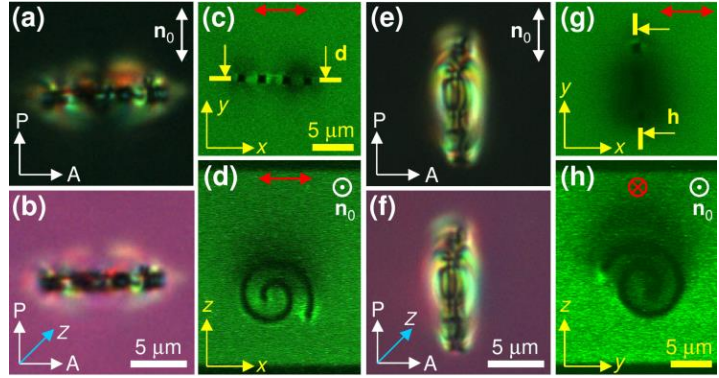

**Fig. 3** Micrographs of singular spiral particles with homeotropic anchoring in a planar nematic liquid crystal cell obtained with (a, b, e and f) polarizing and (c, d, g and h) 3PEF-PM optical microscopies.

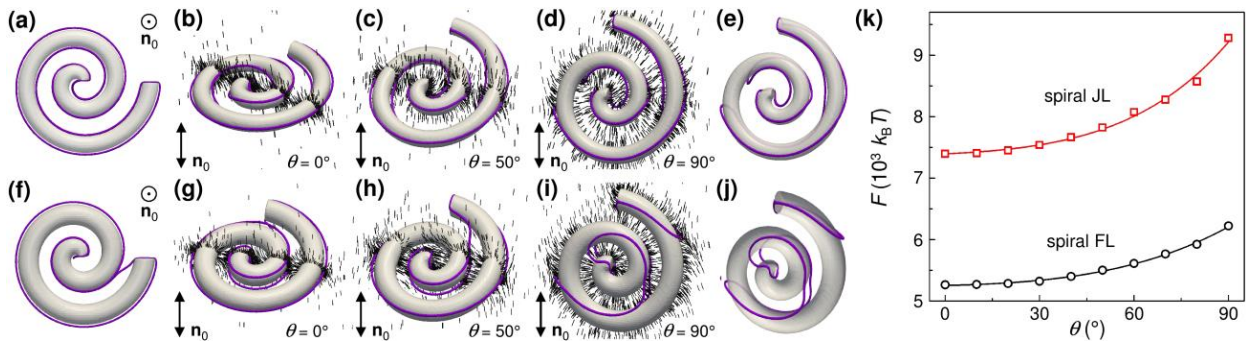

**Fig. 4** (a–j) Numerically calculated models showing (a and f) two ways of wrapping a homeotropic spiral particle with a disclination line $k = -1/2$ and (b–e and g–j) corresponding $\mathbf{n}(\mathbf{r})$ (short black lines) depending on the orientation of spiral particles with respect to $\mathbf{n}_0$. (e and j) Models showing spirals at $\mathbf{a} \perp \mathbf{n}_0$ as in (d) and (i) wrapped only with disclination loops. (k) Landau–de Gennes free energy *vs.* the angle between a normal a to the spiral plane and $\mathbf{n}_0$; solid lines are a guide to the eyes.

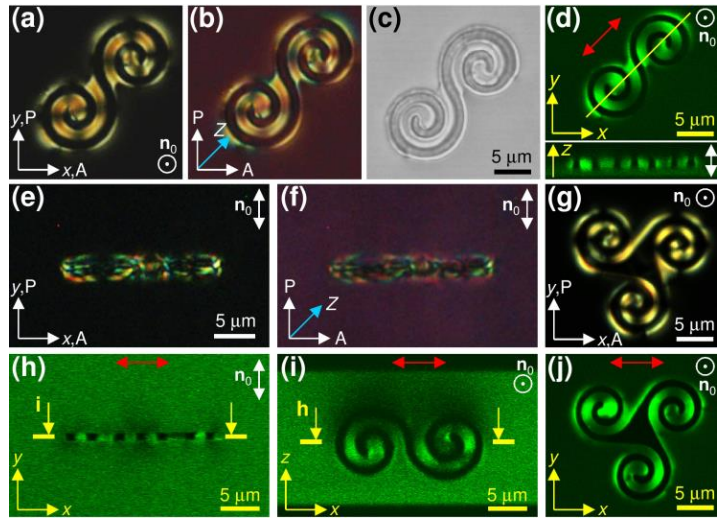

**Fig. 5** Double and triple spiral particles with homeotropic anchoring obtained with (a, b and e–g) polarizing, (c) bright field and (d, h and i) 3PEF-PM optical microscopies: (a–d) double and (g and j) triple spirals in a homeotropic nematic cells. A top image in (d) is an in-plane *xy* view and a bottom image is a cross-sectional view obtained along the bright yellow line in the top part; a double white arrow shows the homeotropic $\mathbf{n}_0$. (e, f, h and i) A double spiral in a planar nematic cell.

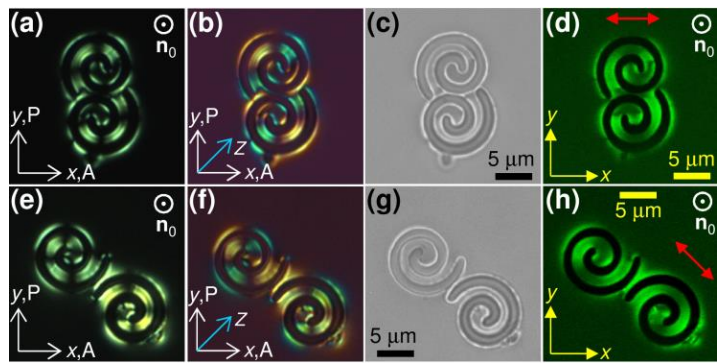

**Fig. 6** Self-assemblies of singular homeotropic spiral particles in a homeotropic nematic liquid crystal obtained with (a, b, e and f) polarizing, (c and g) bright field and (d and h) 3PEF-PM optical microscopies.

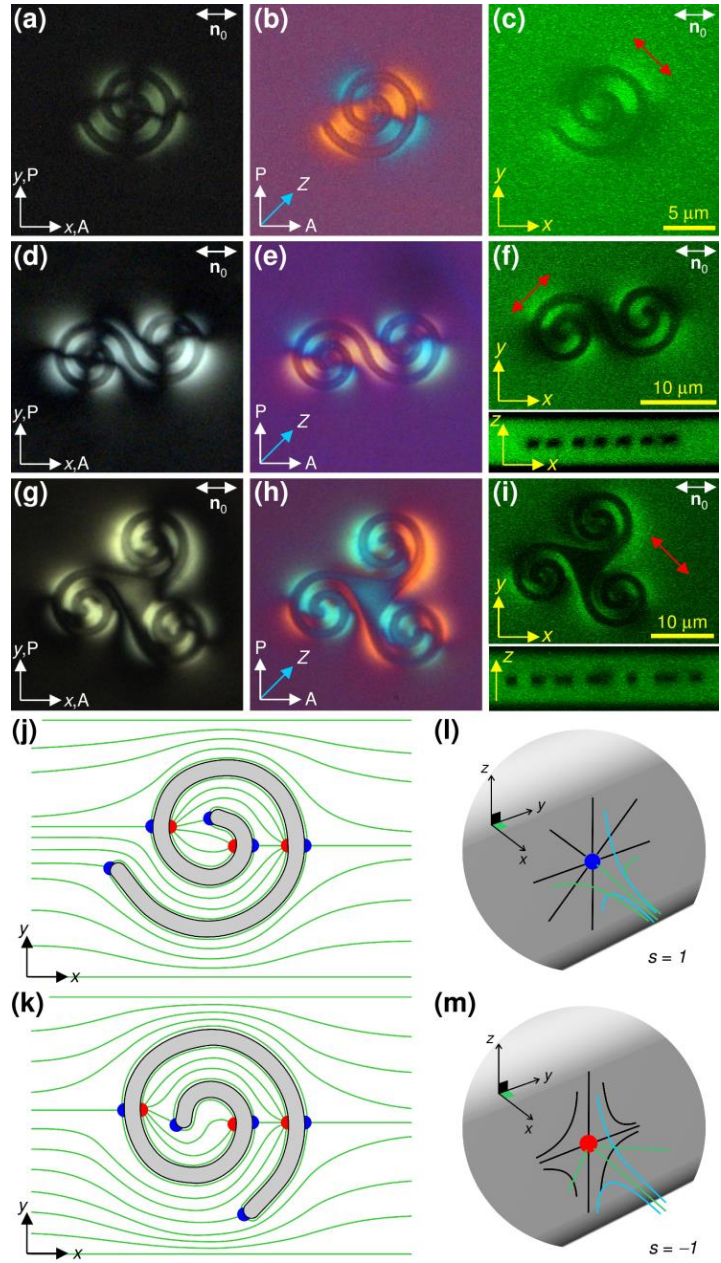

**Fig. 7** Spiral particles with tangential anchoring in a planar nematic cells: (a, b, d, e, g and h) polarizing and (c, f and i) 3PEF-PM optical microscopy micrographs. (j and k) Schematic diagrams of **n**(**r**) around singular spirals having different azimuthal orientations with respect to **n**$_0$. Red and blue filled semicircles show surface point defects boojums. (l and m) Detailed schematic diagrams of **n**(**r**) around boojums.

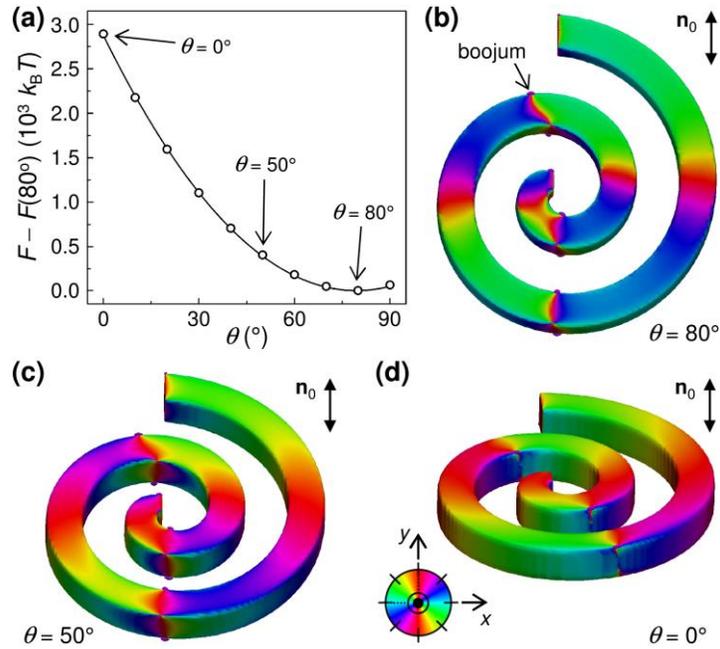

**Fig. 8** Numerical modeling of spiral-shaped nematic colloidal particles with tangential anchoring in a planar nematic cell: (a) Landau–de Gennes free energy vs. an angle y between a normal *a* to the spiral plane and $\mathbf{n}_0$. (b–d) Corresponding $\mathbf{n}(\mathbf{r})$ depending on the orientation of spiral particles with respect to $\mathbf{n}_0$: in-plane azimuthal orientation of $\mathbf{n}(\mathbf{r})$ at the surface of the spiral is shown by color corresponding to a color map in (d). Note that the core of boojums can split to half-integer disclination semiloops.[67]